\documentclass[useAMS,usenatbib]{mn2e}
\usepackage{graphicx,lscape,amssymb}
\usepackage{natbib}
\usepackage{fancyhdr}
\usepackage{subfigure}
\usepackage{amsmath}
\usepackage{mathtools}
\usepackage{tabulary}
\usepackage{color}
\usepackage{booktabs, dcolumn}

\newcommand{\Teff}{$T_{\mathrm{eff}}$}
\newcommand{\Msun}{$\mathrm{M_{\odot}}$}
\newcommand*{\myalign}[2]{\multicolumn{1}{#1}{#2}}

\title{Trace hydrogen in helium atmosphere white dwarfs as a possible signature of water accretion}

\author[Gentile Fusillo et al.]{Nicola Pietro Gentile Fusillo$^1$, Boris T. G\"ansicke$^1$, Jay Farihi$^2$, Detlev Koester$^3$, 
\newauthor Matthias R. Schreiber$^4$, Anna F. Pala$^1$\\$^1$ Department of Physics, University of Warwick, Coventry, CV4 7AL, UK\\$2$ Department of Physics and Astronomy, University College London, London, NW1 2PS, UK\\$3$ Institut f\"ur Theoretische Physik und Astrophysik, University of Kiel, 24098 Kiel, Germany\\$4$ Institute of Physics and Astronomy, Universidad de Valparaiso, Av. Gran Bretana 1111, Valparaiso, Chile\\}

\begin{document}
\maketitle

\label{firstpage}

\begin{abstract}
A handful of white dwarfs with helium-dominated atmospheres contain exceptionally large masses of hydrogen in their convection zones, with the metal-polluted white dwarf GD\,16 being one of the earliest recognised examples.
We report the discovery of a similar star: the white dwarf coincidentally named GD\,17.
We obtained medium-resolution spectroscopy of both GD\,16 and GD\,17 and calculated  abundances and accretion rates of photospheric H, Mg, Ca, Ti, Fe and Ni. The metal abundance ratios indicate that the two stars recently accreted debris which is Mg-poor compared to the composition of bulk Earth. However, unlike the metal pollutants, H never diffuses out of the atmosphere of white dwarfs and we propose that the exceptionally high atmospheric H content of GD\,16 and GD\,17 (2.2$\times 10^{24}$g and 2.9$\times 10^{24}$g respectively) could result from previous accretion of water bearing planetesimals. Comparing the detection of trace H and metal pollution among 729  helium atmosphere white dwarfs, we find that the presence of H is nearly twice as common in metal-polluted white dwarfs compared to their metal-free counterparts. This highly significant correlation indicates that, over the cooling age of the white dwarfs, at least some fraction of the H detected in many He atmospheres (including GD\,16 and GD\,17) is accreted alongside metal pollutants, where the most plausible source is water.  In this scenario, water must be common in systems with rocky planetesimals.
\end{abstract}


\section{Introduction}
Because of the high surface gravities (log $g$) of white dwarfs,  heavy elements sink out of their photosphere on timescales much shorter than their cooling ages \citep{alcock+Illarionov80-1, paquetteetal86-1, dupuisetal92-1, koesteretal97-1}. This process of gravitational settling leads to stratified atmospheres with outer layers purely composed of the lightest elements, H or He. However 25 to 50 percent of all white dwarfs  also show photospheric contamination by trace metal species  (\citealt{zuckermanetal03-1};  \citealt{zuckermanetal10-1}; \citealt{koesteretal14-1}; \citealt{barstowetal14-1}). The now widely  accepted explanation for the origin of these metals is accretion of tidally disrupted planetesimals \citep{jura03-1, debesetal12-1, verasetal14-2,vanderburgetal15-1}.

White dwarf classification is purely based on spectroscopic appearance with the two main classes being spectra dominated by Balmer lines (called DA) and those dominated by He I absorption (called DB). In most white dwarfs the element dominating the spectroscopic appearance (H or He) also corresponds to the main component of the star's atmosphere.
In both magnitude limited and volume limited samples $\simeq80$ per cent of white dwarfs have H atmospheres with the remaining $\simeq20$ per cent having He ones \citep{giammicheleetal12-1, Kleinmanetal13-1}.

Several detailed spectroscopic studies have shown that a significant fraction of He atmosphere white dwarfs also show H$\alpha$ absorption \citep{vossetal07-1, bergeronetal11-1, koesteretal15-1}. In this article we refer to these stars as He+H white dwarfs or He+H+Z white dwarfs, if also metals are detected. Similarly, stars with no detectable H are referred to as He and He+Z white dwarfs. 
In contrast to metal contamination, the origin of H in He-atmosphere white dwarfs is still under discussion. 
Most recently \citeauthor{koesteretal15-1} (\citeyear{koesteretal15-1}, from here on KK15) carried out a comprehensive analysis of 1107 He white dwarfs, ultimately stating that the atmospheric composition of He+H white dwarfs is most likely the result of convective mixing of a primordial H layer. In other words, KK15 argued  that He+H white dwarfs begin their life with relatively thin H layers which get progressively diluted by the underlying He envelope as the stars cool and develop increasingly deep convection zones.

A handful of He-atmosphere white dwarfs are known to have exceptionally high H abundance. One outstanding example is the metal polluted white dwarf GD\,16 (also known as HS0146+1847, \citealt{koesteretal05-1}) which, despite showing a spectrum dominated by H absorption, has actually a He atmosphere  with a large H component and metal contamination. To date, only three similar white dwarfs have been discovered, all of which are also metal polluted (GD\,362, PG\,1225$-$079, and SDSS\,J124231.07+522626.6, see \citealt{gianninasetal04-1, kawka+vennes05-1}, \citealt{kilkenny86-1}, and \citealt{raddietal15-1}).
Here we report new spectroscopy of GD\,16, and the discovery of an additional white dwarf with similar properties and name,  GD\,17 (also known as SDSS\,J014934.45+240046.75). We present a detailed study of the atmospheric composition of GD\,16 and  GD\,17, and discuss the nature of the accreted material. 
\citet{farihietal13-1} and \citet{raddietal15-1} showed that the excess abundance of accreted O in GD\,61 and SDSS\,J1242+5226 could be linked to ongoing (or recent) accretion of water-bearing planetary debris which would also explain the exceptional H content. However H never diffuses down in the atmosphere of white dwarfs, and could, therefore, be a detectable record of past accretion episodes of water-bearing planetesimals long after the metals associated with this event have diffused out of sight.
In order to explore this hypothesis we analyse the sample of He white dwarfs examined by KK15 exploring the correlation between the presence of trace H and metal pollution. Independent on whether the currently detected accretion event is dry (e.g. \citealt{kleinetal11-1, dufouretal12-1})
or wet (e.g. \citealt{farihietal13-1,  raddietal15-1}), the detection of metals is an unambiguous signpost for the presence of planetary systems at these white dwarfs.
We find that the incidence of H in metal polluted white dwarfs (i.e. He+H+Z) is far higher than in their  metal-free counterparts (He+H).
This statistically significant result, combined with the unusual composition of GD\,16 and GD\,17  strongly suggests that a significant number of He+H+Z white dwarfs  have acquired at least some of their atmospheric H via accretion of water-bearing planetary remnants.

\section{The peculiar white dwarfs GD\,16 and and GD\,17}

\begin{figure*}
\includegraphics[width=0.85\textwidth]{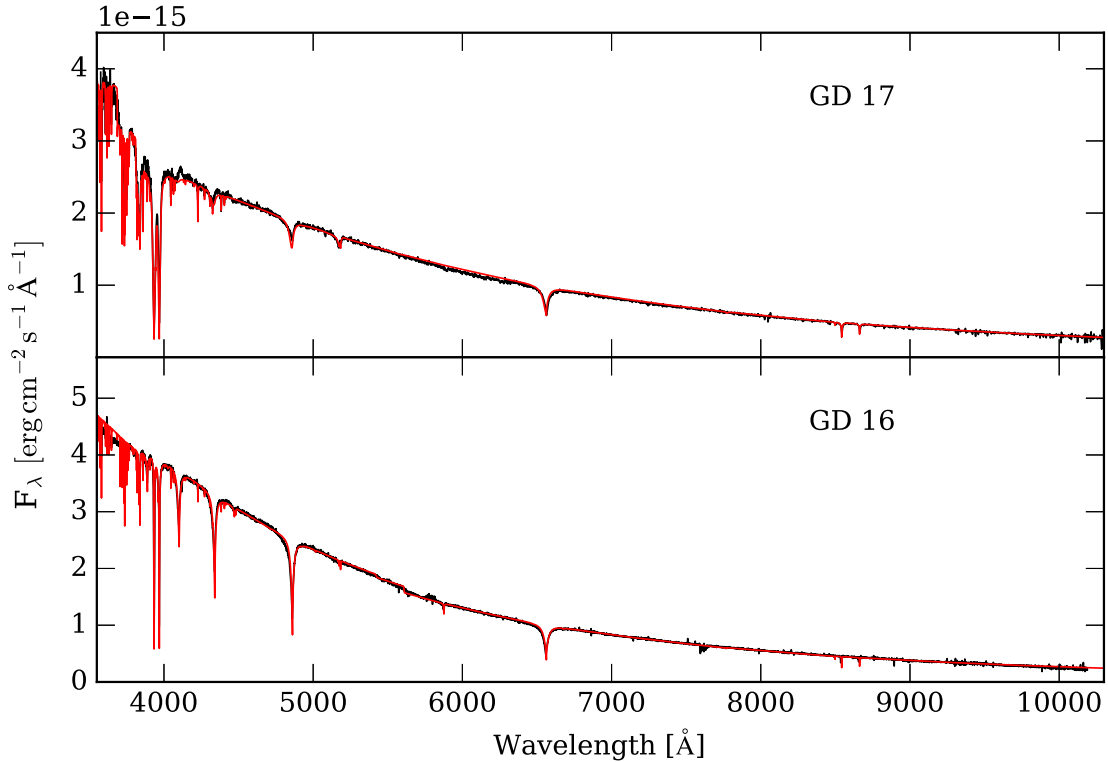}
\caption{\label{gd17_sdss} SDSS spectrum of GD\,17 and VLT X-shooter spectrum of GD\,16 (black). Model spectra using the atmospheric parameters listed in Table\,\ref{GD16_abb} are shown in red.}
\end{figure*}

\subsection{Observations}
GD17 was first identified from the inspection of its SDSS spectrum (Fig.\,\ref{gd17_sdss}) by \citet{gentilefusilloetal15-1} and independently by \citet{kepleretal15-1}. The SDSS spectrum does not contain any He I absorption, but the asymmetric shape of H$\beta$\- suggests an unusual atmospheric composition. 
We obtained additional optical spectra of GD\,17 on 2013 September 4 at the MMT Observatory on Mt. Hopkins (AZ).  The star was observed using the Blue Channel Spectrograph facility instrument and the Echellette grating, providing  simultaneous coverage over 3100$-$7200\,\AA.  
In practice, the bluest order has a blaze function that is difficult to correct, and therefore the shortest usable wavelength range is 3200$-$3300\,\AA .  GD\,17 was placed on a $1''$ slit for seven integrations of 1200\,s each, yielding a total on-source time of 140\,min. A resolving power of $R\approx8200$ was measured from the full width at half maximum of the  O\,{\sc i} 5577\,\AA \ sky line.  The spectrophotometric standard star BD+28\,4211 was observed in a similar manner, as were HeNeAr comparison lamps for wavelength calibration.

The data were reduced manually using the {\sc echelle} package within {\sc iraf}.  A bias frame was subtracted from each exposure, and individual, order-separated spectra were extracted from each resulting frame.  No correction was made for scattered light, nor was flatfielding performed; the available lamps produce insufficient signal in the bluest three orders.  The bright standard was extracted and used as a reference to extract the arc lamp and science target, as the latter were difficult to trace in the highest and lowest orders, especially near their edges. 
The science target data were extracted using a median sky subtraction, variance weighting, and a cleaning algorithm for outliers (e.g.\ cosmic rays).  The extracted spectral orders for GD\,17 were wavelength calibrated using the comparison lamp data, and relative flux calibration was performed by comparing the standard star observations with the spectrophotometric database, and thereby removing the instrument response.  
A final spectrum for each echellette order was created via a signal-to-noise ratio (S/N) weighted average of individual spectra. A combined  spectrum was obtained joining all orders together and averaging the flux in the overlapping regions. 

New optical observations of GD16 were obtained at the VLT observatory on 2014 October 20 with the X-shooter spectrograph \citep{vernetetal11-1} in stare mode, in photometric conditions using a $1''$ slit aperture for the UVB arm and $0.9''$ for the VIS arm. We acquired a single exposure of 1221\,s and 1255\,s for the UVB and VIS arm respectively (Fig.\,\ref{gd17_sdss}).
Using a sky line at 8060\,\AA  \- we measure a full width at half maximum of 1.01\,\AA \- corresponding to a resolving power of $R\approx7980$ at this wavelength.
The spectra were reduced using the standard procedures within the {\sc reflex}\footnote{http://www.eso.org/sci/software/reflex/} reduction tool developed by ESO. The spectral data in the NIR arm had S/N $<$ 5 and was thus insufficient for analysis.

\subsection{Atmospheric parameters}
\subsubsection{GD\,16}
\label{spec_an}
\begin{figure*}
\subfigure{\includegraphics[width=0.45\textwidth]{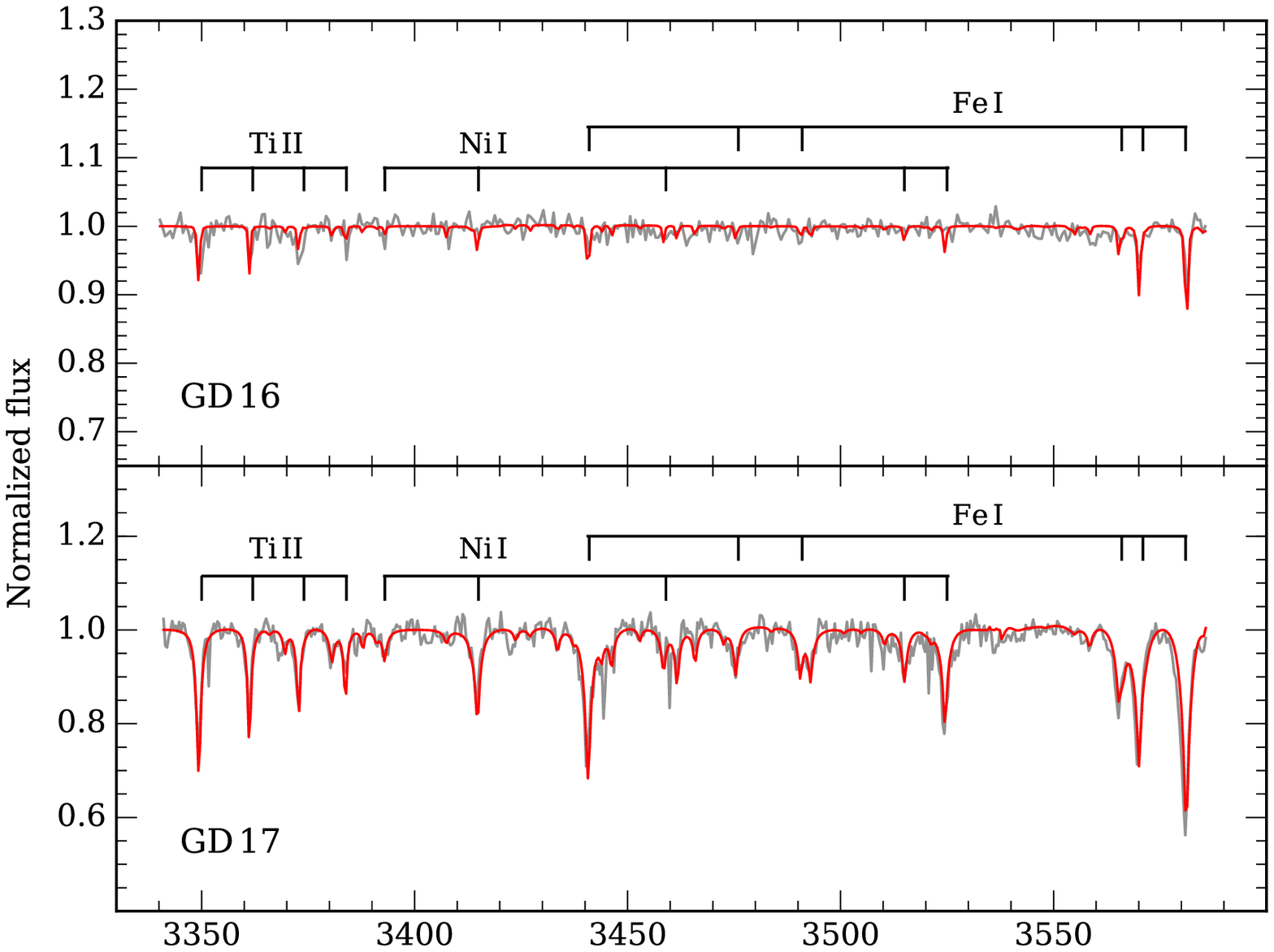}}
\subfigure{\includegraphics[width=0.45\textwidth]{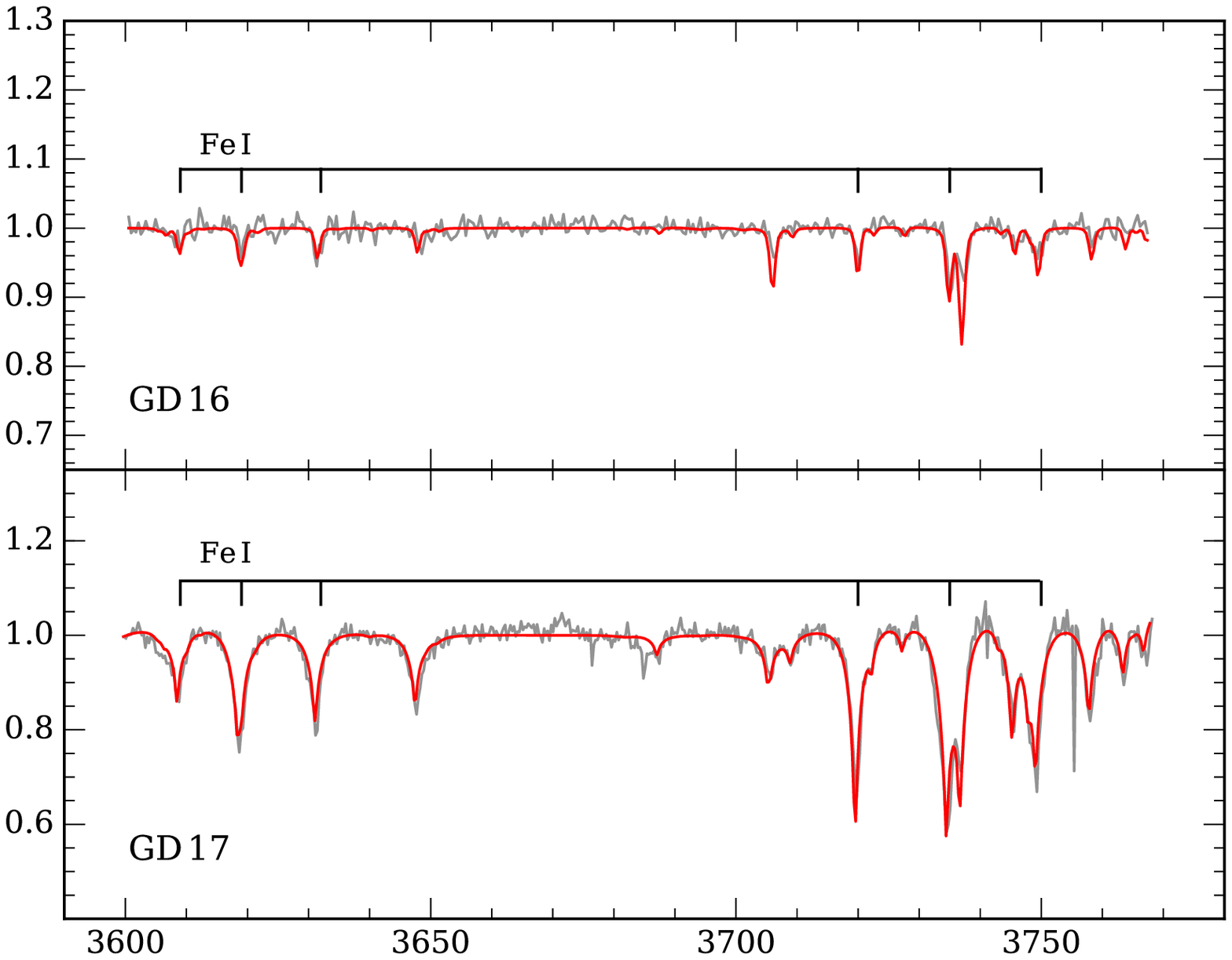}}
\subfigure{\includegraphics[width=0.45\textwidth]{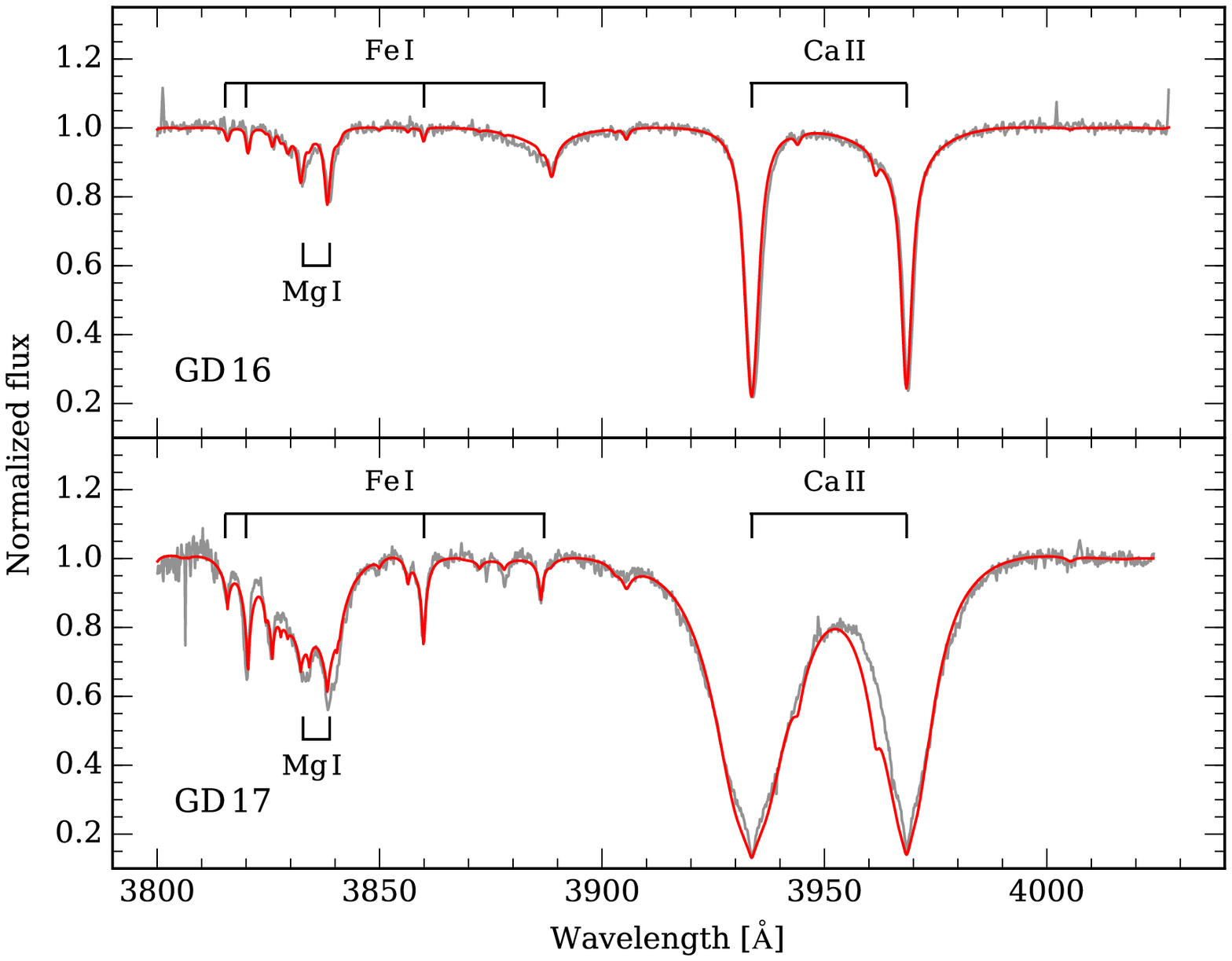}}
\subfigure{\includegraphics[width=0.45\textwidth]{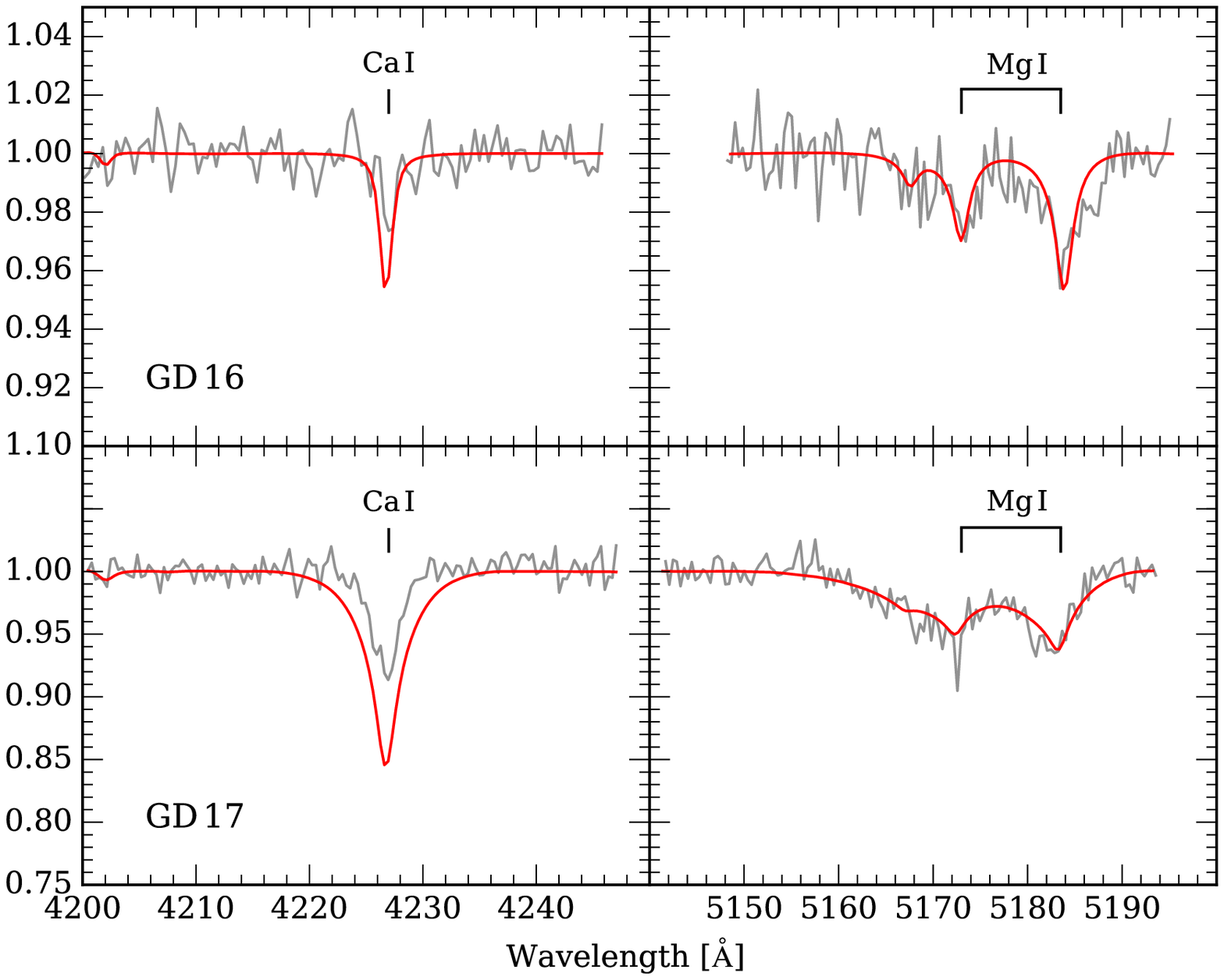}}
\caption{\label{zooms} Relevant absorption features in the X-shooter spectrum of GD\,16 (top panels) and MMT spectrum of GD\,17 (bottom panels). Best-fitting model spectra (Table\,\ref{GD16_abb}) are overplotted in red. As described in section \ref{spec_an}
our models do not correctly reproduce the strength and profile of the 4227\,\AA \- Ca\,I \- line.}
\end{figure*}

\citet{koesteretal05-1} found that no satisfactory fit of the Balmer lines could be achieved using pure H atmosphere models, which led to the realization that GD\,16 has a He dominated atmosphere. In cool He-rich white dwarfs line broadening of H, He and metal lines is dominated by the interactions with neutral He and surface gravity cannot be measured. Adopting a canonical log\,$g$ = 8.0, \citet{koesteretal05-1} determined an effective temperature of $11\,500 \pm 350$\,K. We adopt these parameters as the starting point for analysing the X-shooter spectrum  of GD\,16. Fixing log\,$g$ = 8.0, we calculated a grid of model atmospheres with \Teff \- and the Ca and H abundances as free parameters. In order to achieve a good fit, we had to adjust the line broadening parameters and assume a particularly weak neutral He to H interaction. This dramatically improves the quality of the fit compared to a pure H grid and the best-fitting model was found for \Teff \,$=11\,000\pm500$\,K. However obtaining a fully satisfactory fit has proven to be extremely challenging and our models do not correctly reproduce the profile and strength of the 5875\,\AA \- and 4472\,\AA \- He\,I  lines and of the 4227\,\AA \- Ca\,I line.
Adopting \Teff \,$=11\,000$\,K, we then determined photospheric abundances for H, Mg, Ca, Ti and Fe and estimated upper limits for O, Al, Si, and Ni (Fig.\,\ref{zooms}; Table\, \ref{GD16_abb}).

\begin{table}
\centering
\caption{\label{GD16_abb} J2000 coordinates and proper motions from the Absolute Proper motions Outside the Plane catalogue (APOP, \citealt{qietal15-1}), SDSS magnitudes, and physical parameters  derived from the VLT/X-shooter spectrum of GD\,16, and the SDSS and MMT spectra of GD\,17. The surface gravity of both stars could not be determined and was fixed to log $g$\,= 8.0  (Sec.\,2.2), consequently we do not estimate uncertainties for the masses and radii of these stars. $q_{\mathrm{cvz}}$ indicates the mass of the convection zone expressed as the logarithmic fraction of the mass of the white dwarfs.}
\begin{tabular}{ll D{?}{\,\pm\,}{5.3} D{?}{\,\pm\,}{5.3}}
\hline
 & & \multicolumn{1}{l}{\textbf{GD\,16}} & \multicolumn{1}{l}{\textbf{GD\,17}}\\
\hline \\[-1.5ex]
\multicolumn{2}{l }{R.A. (J2000)} & \multicolumn{1}{r }{01$^\mathrm{h}$ 48$^\mathrm{m}$ 56.81$^\mathrm{s}$} & \multicolumn{1}{r }{01$^\mathrm{h}$ 49$^\mathrm{m}$ 34.38$^\mathrm{s}$}\\
\multicolumn{2}{l }{Dec (J000)} & \multicolumn{1}{r }{19$^\mathrm{\circ}$ 02$'$ 27.60$''$} & \multicolumn{1}{r }{24$^\mathrm{\circ}$ 00$'$ 46.55$''$}\\ 
\multicolumn{2}{l }{$u$ (AB mag)} & 15.55?0.02 & 16.24?0.02\\
\multicolumn{2}{l }{$g$ (AB mag)} & 15.51?0.02 & 15.99?0.01\\
\multicolumn{2}{l }{$r$ (AB mag)} & 15.71?0.01 & 16.07?0.01\\
\multicolumn{2}{l }{$i$ (AB mag)} & 15.85?0.01 & 16.16?0.01\\
\multicolumn{2}{l }{$z$ (AB mag)} & 16.03?0.02 & 16.32?0.01\\
\multicolumn{2}{l }{$\mu$\,R.A. (mas/yr)} & 236.2?16.2& 125.7?13.6\\
\multicolumn{2}{l }{$\mu$\,Dec (mas/yr)} & -100.8?9.7 & 22.3?16.7\\
\hline\\[-1.5ex]
\multicolumn{2}{l }{\Teff (K)} & 11\,000?500 & 8,300?500\\
\multicolumn{2}{ l }{fixed log $g$} & 8.0 & 8.0\\
\multicolumn{2}{ l }{Mass (M$_\odot$)} & 0.583 & 0.579\\ 
\multicolumn{2}{ l }{Radius (R$_\odot$)} & 0.013 & 0.013\\ 
\multicolumn{2}{ l }{Cooling age (Gyr)}  & 0.51 & 1.08\\
\multicolumn{2}{ l }{$q_{\mathrm{cvz}}$}  &-5.3 & -5.2\\[.5ex]
\hline
\end{tabular}
\end{table}

\subsubsection{GD 17}

In order to determine the effective temperature of GD\,17 we relied on the available SDSS spectrum. Assuming log\,$g$ = 8 and accounting for the uncertainty in the SDSS calibration, the slope of the spectrum fixes \Teff \- at  $8300 \pm 500$\,K. We derived photospheric abundances for H, Mg, Ca, Ti, Fe, and Ni and estimated upper limits for Al and Si from the MMT spectrum, and for O from the SDSS spectrum (using OI lines at 7774.081, 7776.300 and 7777.527$\,\AA$, Fig.\,\ref{zooms}; Table\, \ref{GD16_abb}), using the procedure and line broadening parameters adopted above for GD\,16. 

\subsection{Accretion rates}
\label{acc_rates}
He-rich white dwarfs cooler than  \Teff \,$\simeq 25\,000$\,K develop deep convection zones \citep{tassouletal90-1, bergeronetal11-1}.
Accreted elements heavier
than H and He will eventually diffuse below these convection zones  on timescales that depend on several factors including the atomic weight and the convection zone depth ($q_{\mathrm{cvz}}$\,=\,log($M_{\mathrm{cvz}}$ /$M_{\mathrm{*}}$), with $M_{\mathrm{cvz}}$ the mass of the convective layer and $M_{\mathrm{*}}$ the mass of the white dwarf). 
For GD\,16 and GD\,17, we calculated $q_{\mathrm{cvz}}$ of $-5.3$ and $-5.2$, respectively as well as the diffusion time-scales and diffusion velocities for each detected element \citep{koester09-1}. Assuming a steady-state accretion (i.e. balance of rate of accretion and diffusion), we then  calculated their accretion rates by dividing the mass of each element by its sinking timescale.
It is important to bear in mind that these stars have deep convection zones and that the calculated accretion rates represent averages over long ($\sim 1$\,Myr) diffusion timescales.
The results of our calculations are shown in Table\,\ref{GD16_mass}.  
Excluding pollutants for which we can only estimate upper limits, we calculated lower limits on the total average accretion rates of 6.2$\times 10^{7}$\,$\mathrm{g\,s^{-1}}$ and 4.7$\times 10^{7}$\,$\mathrm{g\,s^{-1}}$ for GD\,16 and GD\,17, respectively. Typical accretion rates of metal-polluted He white dwarfs with similar \Teff range between $10^{7}$\,$\mathrm{g\,s^{-1}}$ and $10^{10}$\,$\mathrm{g\,s^{-1}}$ , so the values calculated for GD\,16 and GD\,17 are relatively modest. \citep{girvenetal12-1, bergforsetal14-1, koesteretal14-1}.

\begin{figure}

\includegraphics[width=\columnwidth]{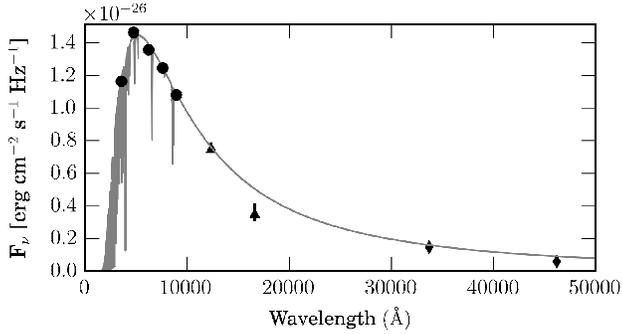}
\caption{\label{gd17_SED} Spectral energy distribution of GD\,17. SDSS $ugriz$, 2MASS $J, K$ and WISE $w1, w2$ magnitudes are plotted as dots, triangles and diamonds respectively. The best-fit model (Table\,\ref{GD16_abb}) is shown in grey. Both near and mid infrared detections are consistent with the flux from the white dwarf alone.}
\end{figure}

In the case of GD\,16 the detection of notable infrared excess \citep{farihietal09-1} strengthens the case for steady state accretion.  However no infrared excess is detected at GD\,17 (Fig.\,\ref{gd17_SED}). This non-detection in consistent with the apparent lack of debris discs around similarly cool metal-polluted white dwarfs observed with \textit{Spitzer} \citep{xuetal12-1, bergforsetal14-1}. The totality of evidence suggests that many such cool stars may still have discs, but they are too tenuous to be detected  \citep{bergforsetal14-1}.
Nonetheless, due to the lack of a detectable infrared excess  at GD\,17, and given the $\sim 1$\,Myr diffusion timescales, it is possible this star is no longer accreting.

\begin{table*}
\newcommand\Tstrut{\rule{0pt}{2.8ex}}         
\caption{\label{GD16_mass} Results of the analysis of the metal pollution of  GD\,16 and of GD\,17 using the atmospheric parameters listed in Table\,\ref{GD16_abb}. From left to right:  photospheric element abundances, the mass of metals in the convection zone, diffusion time-scales, and accretion rates assuming steady-state accretion (see \citealt{koester09-1}, for more details). }
\begin{tabular}{lD{?}{\,\pm\,}{5.3}rcrl|@{\hspace{2em}}D{?}{\,\pm\,}{5.3}rcr}
\hline
&\multicolumn{4}{c}{\textbf{GD\,16}}&&\multicolumn{4}{c}{\textbf{GD\,17}}\\
\hline
Element &\multicolumn{1}{c}{[Z/He]}& \myalign{c}{$M_{\mathrm{Z}}$} & \myalign{c}{$\tau$} & \myalign{c}{$\dot{M}_{\mathrm{Z}}$}& &\multicolumn{1}{c}{[Z/He]}& \myalign{c}{$M_{\mathrm{Z}}$} & \myalign{c}{$\tau$} & \myalign{c}{$\dot{M}_{\mathrm{Z}}$}\Tstrut\\
& & ($10^{22}$\,g)& Myr & $10^{7}$\,$\mathrm{g/s}$&& & ($10^{22}$\,g)&  Myr & $10^{7}$\,$\mathrm{g/s}$\\
\hline
H & -2.80?0.20 & 216.20 &-&\multicolumn{1}{c}{-}&& -2.80?0.50& 288.01 &-&\multicolumn{1}{c}{-}\Tstrut\\
O &\multicolumn{1}{l}{$<-6.00$}& $<2.16$  & 2.64 & $<26.15$&&\multicolumn{1}{l}{$<-4.70$*}&$<57.50	$  & 4.70 & $<336.71$\\
Mg &-7.40?0.10 &  0.13 & 2.85 & 1.46 && -7.50?0.10 &0.14 & 4.02 & 1.10\\
Al & \multicolumn{1}{l}{$<-8.20$} &$<0.02$ & 2.11 & $<0.35$&& \multicolumn{1}{l}{$<-7.20$}& $<0.08$ & 3.80 & $<0.65$\\
Si & \multicolumn{1}{l}{$<-7.20$} &$<0.24$  & 1.73 & $<4.41$&& \multicolumn{1}{l}{$<-7.20$}& $<0.32$ & 3.32 & $<3.06$\\
Ca & -8.65?0.10 & 0.01 & 1.15 & 0.34&&-8.75?0.10& 0.01 & 1.54 & 0.27\\
Ti & -10.40?0.20 & 0.0003 & 1.10 & 0.01& & -10.20?0.20&0.0005 & 1.44 & 0.01\\
Fe & -7.64?0.15 & 0.17 & 1.26 & 4.36& & -7.80?0.10&0.16 & 1.64 & 3.09\\
Ni & \multicolumn{1}{l}{$<-8.80$} & $<0.01$ & 1.37 & $<0.29$&& -9.00?0.20& 0.01 & 1.78 & 0.20\\
\hline

Sum**& &0.31 & &  6.17&& & 0.32 & &  4.67\\
\hline
\multicolumn{8}{l}{\emph{Note}.*Values obtained from SDSS spectrum}\\
\multicolumn{8}{l}{\hspace{0.62cm} **Upper limits are not included}
\end{tabular}
\end{table*}

\subsection{Metal accretion and parent body composition}
\begin{figure}
\hspace{-0.26cm}\includegraphics[width=1.05\columnwidth]{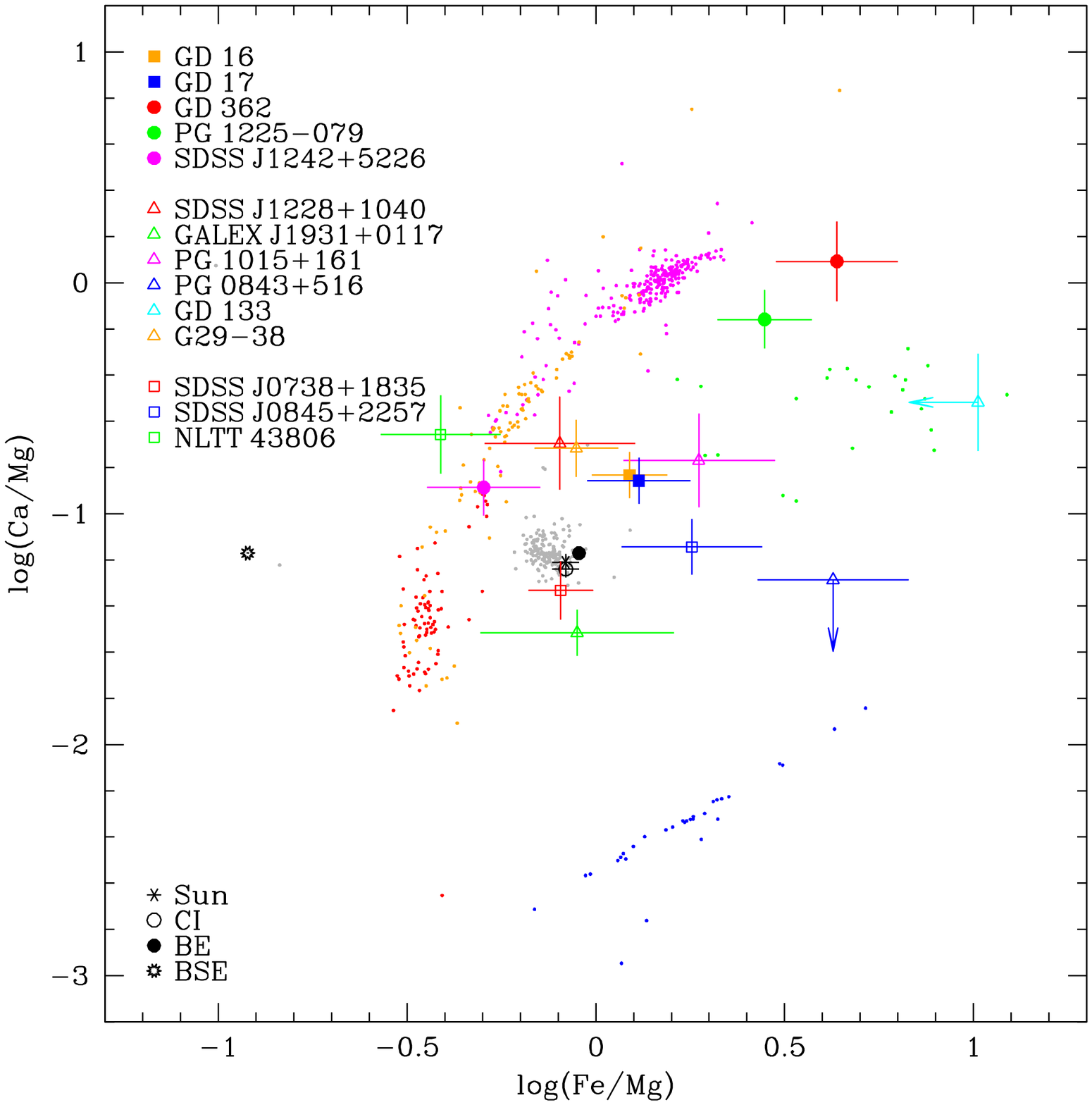}
\caption{\label{m_ratios} Ca/Mg and Fe/Mg logarithmic number ratios of GD\,16 and GD\,17, compared to those of bulk Earth and bulk silicate Earth (BE and BSE; \citealt{mcdonough00-1}), solar abundances and CI chondrites (Sun and CI; \citealt{lodders03-1}), and different meteorite classes (grey=carbonaceous chondrites, green = mesosiderites, blue = pallasites, red = diogenites, orange = howardites, magenta = eucrites; \citealt{nittleretal04-1}). Also shown are the abundance ratios for the white dwarfs GD\,362 \citep{xuetal13-1}, PG\,1225--079 \citep{xuetal13-1}, and SDSS\,J1242+5226 \citep{raddietal15-1}, as well as several other well-studied He-atmosphere \citep{zuckermanetal11-1,dufouretal12-1,wilsonetal15-1} and H-atmosphere white dwarfs \citep{gaensickeetal12-1,xuetal14-1}.}
\end{figure}

In addition to both having He atmospheres, but spectra dominated by Balmer lines, GD\,16 and GD\,17 also show a  similar degree and type of metal pollution.
Abundance ratios inferred for the accreted debris indicate that both stars are polluted by material which appears Mg depleted relative to the composition of  bulk Earth. We calculated  
$\mathrm{Ca/Mg}\simeq0.14$ for GD\,16 and $\mathrm{Ca/Mg}\simeq0.15$ for GD\,17; while  for bulk Earth $\mathrm{Ca/Mg}=0.07$  \citep{mcdonough00-1}. 
Additionally, we find $\mathrm{Fe/Mg}=1.3$ and $\mathrm{Fe/Mg}=1.2$ for GD\,16 and GD\,17 respectively, while for bulk Earth $\mathrm{Fe/Mg}\simeq 0.9$ \citep{mcdonough00-1}. In Fig.\,\ref{m_ratios} we compare these ratios with those of solar system meteorites and other well studied metal polluted white dwarfs. Our estimates suggest that the parent bodies accreted by GD\,16 and GD\,17 had peculiar compositions that do not match any of the major metorite families in the solar system. 
However, we know that similar Mg poor compositions have been measured for a few other known metal polluted white dwarfs (Fig.\,\ref{m_ratios}). 

\begin{figure*}
\includegraphics[width=0.8\textwidth]{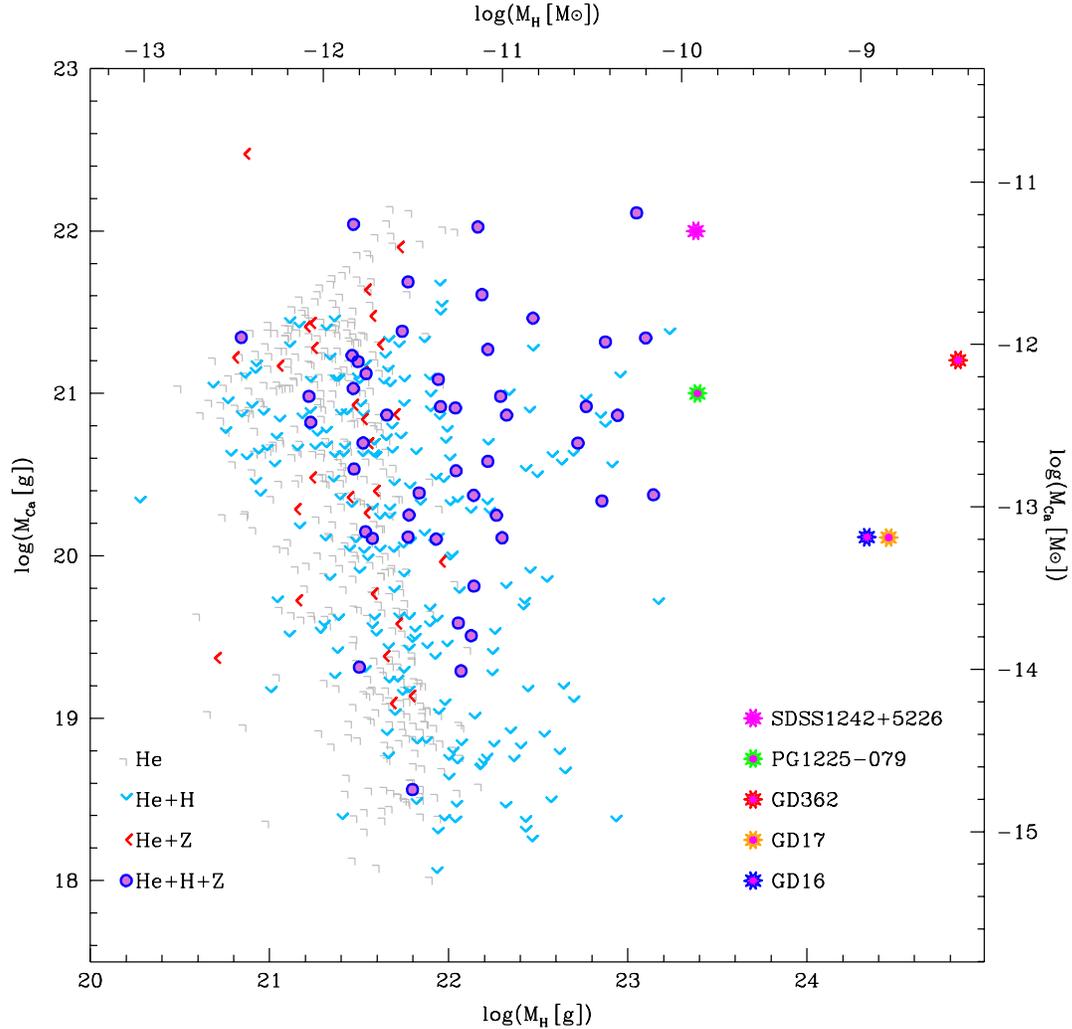}
\caption{\label{H_boris_plot} Distribution of total H and Ca  masses in the convention zones for the white dwarfs in the sample of KK15 with \Teff\,$\leq 17\,000$\,K, plus the particularly H-rich stars GD\,16, GD\,17, GD\,362, PG\,1225$-$079 and SDSS\,J1242+5226. 
For He, He+H and He+Z white dwarfs upper limits on $M_{\mathrm{H}}$ and $M_{\mathrm{Ca}}$, $M_{\mathrm{Ca}}$, and $M_{\mathrm{H}}$ are shown, respectively.} 
\end{figure*}

We note that we assumed steady state accretion in our calculations , but in Section\,\ref{acc_rates} we cautioned that GD\,17 may no longer be accreting. In a declining phase of accretion, because of the short settling timescales, the abundances of Ca and Fe decrease faster than that of Mg (Table\,\ref{GD16_mass}). Consequently the current atmospheric pollutant of GD\,17 could have originated from a parent body with even higher Ca/Mg and Fe/Mg ratios than those inferred from steady state accretion.
Simulations have shown that Mg-depleted rocky planetesimals can form in planetary systems \citep{jadeetal12-1} and GD\,16 and GD\,17 may be examples of this.

\subsection{The origin of atmospheric H}
\label{atm_H} The most distinctive aspect of GD\,16 and GD\,17 is their unusually high H abundance and the consequent DA spectral appearance despite their He atmospheres. Spectral modelling reveal the H-mass in the convection zone is $M_{\mathrm{H}}= 2.2$\,$\times 10^{24}$g  for GD\,16 and $M_{\mathrm{H}}=2.9$\,$\times 10^{24}$g for GD\,17, two of the highest values among all known He atmosphere white dwarfs (\citealt{raddietal15-1}; KK15).
To date only three other similar stars have been discovered, all of which are also heavily metal-polluted (SDSS\,J1242+5226 \citealt{raddietal15-1}, GD\,362 \citealt{gianninasetal04-1, kawka+vennes05-1}, PG\,1225$-$079 \citealt{kilkenny86-1}), although their cool He-atmospheres lead to easily detectable Ca lines. Three of these stars (GD\,16,  GD\,362 and PG1225$-$079) also have dusty debris discs detected in the infrared \citep{becklinetal05-1,kilicetal05-1,farihi09-1}.
The fact that the five He atmosphere white dwarfs with the highest H content known to date are all showing the signature of the ongoing  or recent accretion of 
rocky planetary remnants invites the speculation that the H in these white dwarfs could have been delivered alongside the metal pollutants. 
This is certainly a very likely scenario for SDSS\,J1242+5226. This star shows O abundance significantly in excess to what is expected for the accretion of rocky material, i.e. common metal oxides. The accretion of water-bearing bodies also provide an explanation for the large H content of this white dwarf \citep{raddietal15-1}. 
In contrast, GD\,362 does not show any detectable O absorption lines, and the upper limits on the O abundance reveal that the debris causing the currently observable metal pollution could not have had a significant water component \citep{xuetal13-1}. 
Similarly neither GD\,16 nor GD\,17  show detectable  O lines in their spectra, but our upper limits on the O abundance for these two stars are less constraining than those found for GD\,362 (Table\,\ref{GD16_mass}).

However H never diffuses out of the atmospheres of white dwarfs, in contrast to metals which  disappear from sight on timescales of $\sim10^6$ yr. Even though the metals currently in the atmosphere of these stars do not contain any water, H may be a relic of past accretion events. \citet{juraetal09-1} already proposed that the unusually abundant H in the the atmosphere of GD\,362, could be the result of accretion of water-bearing debris. \citet{farihietal10-1} further extended this idea suggesting that, the pattern of H abundances in cool metal polluted white dwarfs is likely a reflection of the diversity of water content in extrasolar planetesimals. In accordance with this hypothesis we speculate that the H in GD\,16 and GD\,17, and potentially in He atmosphere white dwarfs in general could be a lasting record of water accretion, independently of the detection of O.

\section{The incidence of H in He white dwarfs}

\subsection{The SDSS He white dwarfs sample}
Without reliable detection of an O excess, the presence of trace H in a single He atmosphere white dwarf (even in exceptional cases like GD\,16 and GD\,17) does not, on its own, represent evidence of accretion of water-bearing debris. In order to assess whether trace H in He atmosphere white dwarfs is an indicator of water accretion it is therefore necessary to look at the overall population of He atmosphere white dwarfs. We closely examined the sample of SDSS He white dwarfs analysed by KK15.
In their study, KK15 excluded spectra with S/N $<10$ and all objects with peculiarities, such as stars with low mass companions, stars with He II lines (DO) and  obviously magnetic stars. These selection criteria defined a sample of 1107 stars. KK15  also identified all He+Z white dwarfs within this sample via detection of the Ca H+K resonance lines (Fig.\,\ref{H_boris_plot}).

\begin{figure}
\includegraphics[width=0.48\textwidth]{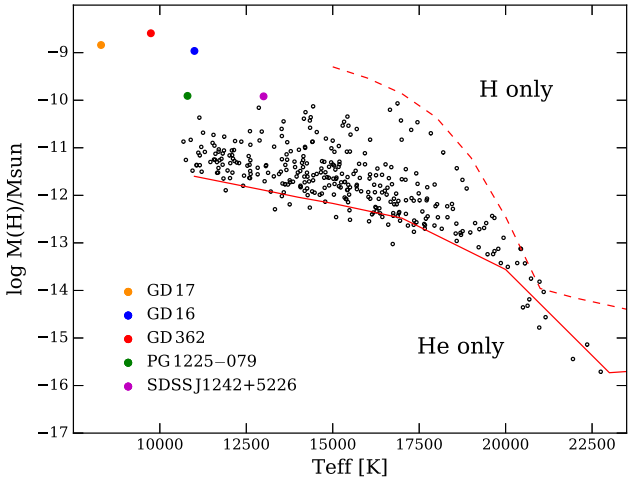}
\caption{\label{H_mass_plot} Distribution of total hydrogen mass in the convection zones of the He+H white dwarfs from  KK15. The continuous red line indicates the H detection limit. The dotted red curve is the expected location for an abundance
log\,[H/He] = $-2$ above which the star would develop a pure H atmosphere.}
\end{figure}

\subsection{Evidence of external H accretion}
\label{incidence}
Balmer lines get weaker with increasing \Teff , therefore KK15 carried out of their statistical analysis of  the incidence of H only on He white dwarfs with \Teff $\leq$\,23\,000\,K (1037 objects).They found that $\simeq32$ per cent of the He white dwarfs show traces of H in their atmosphere. When only  the highest S/N spectra are considered (i.e. a total sample of 64 white dwarfs with S/N $\geq 40$) this ratio becomes as high as $\simeq75$, suggesting that the difference between He and He+H white dwarfs is possibly just a question of the quality of the observations, and that all He white dwarfs may have some H.
KK15 argued against accretion from the ISM as the origin of H in He+H white dwarfs by showing that: one, there is no evidence for growing H mass with longer cooling ages, and two, the frequency of He+H white dwarfs is independent of the distance from the Galactic plane. This is in agreement with other previous studies that similarly opposed ISM accretion as the most likely H delivery mechanism \citep{farihietal10-1,bergeronetal11-1}. KK15 concluded that the H present in He+H white dwarfs is most likely primordial and accretion plays no significant role. In other words KK15 suggested that He+H white dwarfs begin their lives with relatively thin H layers which become diluted by the underlying He envelope as the stars cool and develop deeper convection zones.

However, from a closer analysis of the He white dwarfs sample considered by KK15, we find some evidence that cannot be explained by the primordial origin of H.
In the primordial H scenario, He+H white dwarfs formed with the same mass of H that we measure today. Indeed,  in order to explain the range of H masses measured for He+H white dwarfs, KK15 concluded that they must form with H layers ranging from the thick H envelopes of 
DAs to ultra thin H-layers with masses as low as  $10^{-17}$\Msun\ and that all stars with $M_{\mathrm{H}}\lesssim 10^{-9}$\Msun\ will eventually evolve into He+H white dwarfs. In other words, as white dwarfs cool they should simply move horizontally from right to left in Fig.\,\ref{H_mass_plot}. However the observed distribution of He+H white dwarfs, in the temperature range where both H and He are detectable (the area in between the red curves in   Fig.\,\ref{H_mass_plot}), is significantly inhomogeneous with most objects clustering close to the lower detection limit (solid red line). Stars with large amounts of H in their He envelopes such as GD16, GD17, and GD362 are clearly exceptional. Under the assumption that all H in these stars is primordial, this distribution implies that the prior stellar evolution of white dwarfs leads preferentially to the formation of thick H layers (and so H-atmosphere white dwarfs), or to thin H layers with masses in the range $M_{\mathrm{H}}\simeq10^{-13}-10^{-11}$\,\Msun, but only rarely to layers of intermediate thickness. KK15 attempted to explain the origin of all He+H white dwarfs with a single mechanism. Here we suggest as an alternative scenario: the existence of an additional mechanism which contributes to the formation of He+H white dwarfs.

About 10$-$12 per cent of the He white dwarfs examined by KK15 are also contaminated by traces of Ca.
Today the standard model for atmospheric metals in white dwarfs is accretion of rocky planetesimals \citep{jura03-1,gaensickeetal08-1,farihietal10-1}. These heavy elements are therefore not related to stellar evolution. 

We investigated the correlation between the presence of H and that of Ca in the He white dwarfs in KK15 sample. 
The detectability of Ca lines drops significantly with increasing \Teff , we therefore limited our analysis to white dwarfs with \Teff $\leq17\,000$\,K (729 objects). 
We found a marked difference in the incidence of H in He+Z white dwarfs compared to metal-free He white dwarfs which becomes apparent when comparing two ratios:

\begin{equation*}
p_{\mathrm{HeH}}\equiv\frac{N_{\mathrm{HeH}}}{N_{\mathrm{HeH}}+N_{\mathrm{He}}}
\end{equation*}

\begin{equation*}
p_{\mathrm{HeHZ}}\equiv\frac{N_{\mathrm{HeHZ}}}{N_{\mathrm{HeHZ}}+N_{\mathrm{HeZ}}}
\end{equation*}

\noindent We find $p_{\mathrm{HeH}}\simeq 32$ per cent (the same value found by KK15 for the entire sample with \Teff $\leq$23\,000\,K), and $p_{\mathrm{HeHZ}}\simeq 61$ per cent.
The standard error for a proportion is given by $\sqrt{p\times(1-p)/n}$ 
where $p$ is the measured proportion and $n$ the number of measurements (i.e. $n=N_\mathrm{HeH}+N_\mathrm{He}$ for $p_\mathrm{HeH}$, and $n=N_\mathrm{HeHZ}+N_\mathrm{HeZ}$ for $p_\mathrm{HeHZ}$). 
For our two independent samples we obtain 
$p_{\mathrm{HeH}} = 0.32\pm0.02$ and 
$p_{\mathrm{HeHZ}} = 0.61\pm0.06$. 
At face value this suggests that the measured difference in the fractions of white dwarfs with H is highly significant.
The average S/N for the metal-free sample and the metal polluted sample are 21.0 and 21.6 respectively and, therefore, any bias caused by  the detectability of H at different S/N would affect $p_{\mathrm{HeH}}$ and $p_{\mathrm{HeHZ}}$ in similar ways. We also find no significant difference in the temperature distributions of the two samples reinforcing that H should be equally detectable for both groups of stars.
We evaluate the  significance of the difference between $p_{\mathrm{HeH}}$ and $p_{\mathrm{HeHZ}}$ by performing the appropriate statistical test for the two independent samples. 
Given the null hypothesis that the two proportions are equal i.e. $p_{\mathrm{HeH}}-p_{\mathrm{HeHZ}}=0$, we can use our best estimate of the total fraction of systems with H
as our best estimate for both samples, i.e.
\begin{equation}
p=\frac{N_{\mathrm{HeHZ}}+N_{\mathrm{HeH}}}{N_{\mathrm{HeH}}+N_{\mathrm{HeHZ}}+N_{\mathrm{He}}+N_{\mathrm{HeZ}}}
\end{equation}
The standard error for the difference of two independent samples is then
\begin{equation}
\sigma_{p_{\mathrm{HeH}}-p_{\mathrm{HeHZ}}}=
\sigma_{p_{\mathrm{HeH}}}-\sigma_{p_{\mathrm{HeHZ}}}
= \nonumber
\end{equation}
\begin{equation}	
\hspace{-3cm} \sqrt{p_{\mathrm{HeH}}\times\left(1-\frac{p_{\mathrm{HeH}}}{N_{\mathrm{HeH}}+N_{\mathrm{He}}}\right)}\nonumber
\end{equation}

\begin{equation}
\hspace{1.8cm}\overline{\rule{0pt}{3.ex}-p_{\mathrm{HeHZ}}\times\left(1-\frac{p_{\mathrm{HeHZ}}}{N_{\mathrm{HeHZ}}+N_{\mathrm{HeZ}}}\right)}
\end{equation}

\noindent
and the corresponding test statistic $z$ is 
\begin{equation}
z=\frac{(p_{\mathrm{HeH}}-p_{\mathrm{HeHZ}})-0}{\sigma_{p_{\mathrm{HeH}}-p_{\mathrm{HeHZ}}}}
\end{equation}
From equation 3 we find $z=5.0$ implying that we can reject the null hypothesis at a  $>5\sigma$ significance level. In other words, under the assumption that H is equally common in metal-polluted He white dwarfs than in metal-free ones, the observed ratios would have less than one in a million chance of occurring. 
In the primordial H or ISM accretion hypothesis, no correlation should exist between the presence of H and accretion of metals (in this case Ca) from an external source.
The most likely explanation for what we observe is therefore that  a significant fraction of the H detected in He+H+Z white dwarfs must have the same origin as the metals, i.e. accretion of planetary debris. 

\section{Discussion}
\label{H_discussion}
Three main theories have been put forward to explain the origin of trace H in He atmosphere white dwarfs: accretion from the ISM \citep{macdonald+vennes91-1, bergeronetal11-1}, convective mixing  of a primordial thin H layer (\citealt{bergeronetal11-1}; KK15) and accretion of water-bearing debris \citep{juraetal09-1,farihietal10-1, verasetal14-1, raddietal15-1}. In their comprehensive analysis of He white dwarfs \citet{bergeronetal11-1} concluded that the primordial H hypothesis cannot explain the entire range of $M_{\mathrm{H}}$ observed in He+H white dwarfs, unless a mechanism of non-complete mixing of H in the convective zone is invoked. They further argued that ISM accretion could also provide a plausible explanation, but this would require H accretion to be effective for some He+H white dwarfs, and not for other stars with similar temperatures and cooling ages.
The more recent work by KK15 showed that no correlation could be found between $M_{\mathrm{H}}$ and either height of the white dwarfs above the galactic plane or their cooling ages, which led KK15 to rule out the ISM accretion theory and strongly favour the primordial H hypothesis. 

In the water accretion scenario, H is delivered onto the white dwarf (as water or bound in hydrated minerals), together with various metal species  as a result of the tidal disruption of asteroids,  comets and possibly even planets \citep{juraetal09-1, farihietal11-1, verasetal14-1, raddietal15-1}. 
Hydrogen can in principle occur in compounds other than water, for example ammonia (NH$_{3}$) and methane (CH$_{4}$) which are also common in our solar system. However, solar system bodies which contain these compounds are also known to harbour far larger amounts of water .
Furthermore, it has become increasingly clear that most metal-polluted white dwarfs accrete volatile-depleted material, with very low mass fractions of C and seemingly devoid of N \citep{jura06-1, farihietal10-1, gaensickeetal12-1, juraetal12-1, xuetal14-2}; with the only noticeable exception of WD\,1425+540 \citep{xuetal17-1}. 
Previous studies examining evidence of water accretion onto white dwarfs relied on detection of O in the atmosphere of these objects, a key diagnostic which can be directly related to ongoing or recent accretion of water (\citealt{juraetal09-1, farihietal10-1, raddietal15-1}, see section \ref{atm_H}). Because of the low resolution and limited S/N of the SDSS spectra, Ca is the only metal detected in most of the He+H+Z white dwarfs in the KK15 sample and, in the absence of O detections, no  direct proof of ongoing or recent water accretion is possible.
However, H never diffuses out of the atmosphere of white dwarfs and could therefore be a lasting record of water accretion long after all metals have diffused out.
For individual white dwarfs it is not possible to claim that the currently observable metal pollution and the trace H result from the accretion of the same body, but the presence of metals undeniably shows that the star hosts a planetary system and is accreting rocky remnants.
Looking at the population of He atmosphere white dwarfs, we  showed that the incidence of H is strongly correlated with the presence of metals (section \ref{incidence}). We demonstrated that white dwarfs which host rocky planetary systems are statistically much more likely to have trace H compared to metal-free white dwarfs.
We therefore propose that the majority of He+H+Z white dwarfs are likely to have accreted at least some fraction of their $M_\mathrm{H}$, at some point during their cooling age, in the form of water or hydrated minerals in rocky planetesimal (e.g. as detected in GD61, \citealt{farihietal13-1}), or comets \citep{verasetal14-1, stoneetal15-1, xuetal17-1}.

\citet{jura+xu12-1} quantitatively estimated the mass fraction of water in the debris accreted by white dwarfs, by comparing the summed mass accretion rates of heavy elements with that of H in a volume-limited sample of He white dwarfs. \citet{jura+xu12-1} concluded that, on average both across the sample and over time, water makes up less than one per cent of the mass of the material accreted by white dwarfs. 
Assuming all H in the atmosphere of the He+H+Z white dwarfs in the KK15 sample is the result of water accretion, the upper limit of accreted water mass for these stars ranges between $6\times 10^{21} \,\mathrm{g}$\- and $1\times 10^{24} \,\mathrm{g}$. Adopting the same statistical approach used by \citet{jura+xu12-1}, on average,  these masses only correspond to  $\sim0.5$\- per cent of the mass of metals accreted by these stars. Therefore our analysis supports that of \citet{jura+xu12-1}. 

One limitation of the analysis presented by \citet{jura+xu12-1} is that their sample was limited to white dwarfs with cooling ages $\lesssim 3.5\times 10^8$\,yr.
Water accretion is most likely not a continuous event:
while large amounts of H may be accumulated at once as the result of the tidal disruption of one large water-rich planetesimal in the solar system (e.g. Ceres), multiple accretion events of small bodies (e.g. small asteroids and comets, \citealt{verasetal14-1}) can gradually contribute to the H mass over the entire cooling age of the white dwarfs. 
This naturally explains why the  outliers with the highest $M_\mathrm{H}$ are all relatively cool and old white dwarfs. 
Applying the analysis of \citet{jura+xu12-1} only to the small sample of cool $M_\mathrm{H}$ outliers, i.e. GD\,16, GD\,17, GD\,362, PG\,1225$-$079 and SDSSJ\,1242+5226 (Tables\,\ref{GD16_abb},\,\ref{GD16_mass}), we find that, on average, the water-to-metal mass fraction in these stars is $\simeq 11$ per cent. Of course applying this type of analysis to only five stars can only provide a weakly constrained upper limit. Nonetheless the water-to-metal mass fraction found is much higher than the $\simeq 0.5$ \- per cent  we calculate for the KK15 sample and from the one per cent found by \citet{jura+xu12-1}, which underlines once more the peculiar nature of these white dwarfs.

Assuming that these five stars accreted H at some point during their cooling age (like we showed is a likely scenario for most He+H+Z white dwarfs), some detail of their accretion history must set them apart from the rest of the white dwarfs in the sample. It is possible that these stars hosted particularly water-rich planetary systems, which, in turn, would also explain why only a handful of these outliers have been discovered to date. Assuming $M_\mathrm{H}$ has a planetary origin, these stars accreted up to $4.6\times 10^{25} \mathrm{g}$ of water, a mass close to the upper limit of the total water content of Earth (estimated to be between 0.06 and 2 per cent by mass,  \citealt{thienenetal07-1, satoetal16-1}). Accretion of the entire $M_\mathrm{H}$ in a single event would imply large and very water rich parent bodies, i.e. entire planets or moons ejected from their orbits \citep{payneetal16-1}.

These peculiar white dwarfs are older than the majority of the other stars examined by both KK15 and \citet{jura+xu10-1}, so, alternatively, their large atmospheric H content could potentially be the result of slow accumulation over their long cooling ages. In this case we would expect more of these H-rich He atmosphere white dwarfs particularly at low temperatures. However once metals have completely diffused out of their atmospheres, the spectrum of these stars would only show H absorption and these white dwarfs could easily not be recognised.
On the other hand, to date, hundreds of metal polluted white dwarfs with \Teff \,$\lesssim 10\,000$\,K (DZs) are known, the vast majority of which have, at most, only minor amounts of  H in their atmospheres \citep{dufouretal07-2}.  
Independently of the specific delivery scenario we propose that  GD\,16, GD\,17 and by analogy GD\,362, PG\,1225$-$079 and SDSS\,J1242+5226 likely owe at least part of their exceptional H content to accretion of water-bearing debris over their cooling age.

\section{Conclusions}
To date four He-atmosphere white dwarfs are known to have exceptionally high amounts of H in their atmosphere, all of which also show significant metal pollution. Here we reported the discovery of a new member of this group, the white dwarf GD\,17  and compared it with the archetypal example of the class, GD\,16.
Our analysis of newly acquired spectroscopy of GD\,16 and GD\,17 revealed that the H contents in the atmosphere of these stars are some of the highest measured among all He atmosphere white dwarfs. 
Unlike metal pollutants H never diffuses out of the atmosphere of a white dwarf and can therefore be a lasting tracer of water accretion. Independently of their specific accretion history we conclude that  GD\,16, GD\,17 and by analogy the other stars in this class (i.e. GD\,362, PG\,1225$-$079 and SDSS\,J1242+5226) likely owe at least part of their exceptional H content to accretion of water-bearing debris. 

We investigated the idea of H delivery via accretion of water-bearing debris by comparing the incidence of H in He and He+Z white dwarfs from the sample analysed by \citet{koesteretal15-1}. We rejected with $>5\sigma$ significance the hypothesis that H is equally common in He and He+Z white dwarfs, and find that trace H is present in 61 per cent of the metal polluted white dwarfs in our sample and in only 32 per cent of the metal-free ones. 
Our analysis indicates that accretion of water-bearing bodies likely represents a significant contribution for the H content of He+H white dwarfs and, though rare compared to dry material, some amount of water must be present in most planetary systems which harbour rocky bodies.

\section*{Acknowledgements}
The research leading to these results has received funding
from the European Research Council under the European
Union’s Seventh Framework Programme (FP/2007-2013) /
ERC Grant Agreement n. 320964 (WDTracer). 
\newline Based on observations made with ESO Telescopes at the Paranal Observatory under programme ID 094.D-0344(A).
\newline NPGF and MRS thank for support from Fondecyt (grant 1141269).

\bibliographystyle{mn_new}

\end{document}